\documentclass[10pt,conference]{IEEEtran}
\usepackage{epsfig,setspace,amsmath,epsf,amssymb,bm,theorem,cite,graphicx, epstopdf,algorithm,float,color,mathtools, authblk}
\usepackage[table,xcdraw]{xcolor}
\usepackage{subcaption}
\usepackage{algpseudocode}

\DeclarePairedDelimiter{\ceil}{\lceil}{\rceil}

\newtheorem{remark}{Remark}

\newcommand{\e}{{\mathbb{E}}}

\IEEEoverridecommandlockouts
\allowdisplaybreaks

\begin{document}

\title{Minimizing the Age of Two Heterogeneous Sources With Packet Drops Via Cyclic Schedulers}

\author[1]{Sahan Liyanaarachchi}
\author[1]{Sennur Ulukus}
\author[2]{Nail Akar}

\affil[1]{\normalsize University of Maryland, College Park, MD, USA}
\affil[2]{\normalsize Bilkent University, Ankara, T\"{u}rkiye}

\maketitle
\let\thefootnote\relax\footnotetext{This work is done when N.~Akar is on sabbatical leave as a visiting professor at University of Maryland, MD, USA, which is supported in part by the Scientific and Technological Research Council of T\"{u}rkiye  (T\"{u}bitak) 2219-International Postdoctoral Research Fellowship Program.}

\begin{abstract}
In a communication setting where multiple sources share a single channel to provide status updates to a remote monitor, source transmissions need to be scheduled appropriately to maintain timely communication between each of the sources and the monitor. We consider age-agnostic scheduling policies which are advantageous due to their simplicity of implementation. Further, we focus on a special class of age-agnostic policies, called \emph{cyclic} schedulers, where each source is scheduled based on a fixed cyclic pattern. We use weighted average age of information (AoI) to quantify the timeliness of communication. We develop a Markov chain formulation to compute the exact mean AoI for the case of two-source cyclic schedulers. Based on the obtained age expression, we develop an algorithm that generates near-optimal cyclic schedulers to minimize the weighted average AoI for two heterogeneous sources, in the presence of channel errors. 
\end{abstract}

\section{Introduction}
Timely communication is necessary in many remote estimation settings, where a source is sampled and the status updates are sent to a remote monitor through a random-delay channel. The timeliness of the status updates is measured through the average AoI of the process given by $\e[\Delta_t]$, where $\Delta_t= t - u(t)$ is the instantaneous age of the source and $u(t)$ is the time of generation of the freshest update available at time $t$ to the remote monitor \cite{yates2020age}. A common scenario that arises in this setting is that multiple sources may share the same channel to provide status updates to the remote monitor \cite{rtsms2019}. We consider the \emph{generate-at-will} model introduced in \cite{lts2015} for status generation where each source is capable of generating  a status update at any time when needed. In this multi-source setting, status updates of some sources may be more important than the others and therefore the weighted average of mean AoI values of the sources (termed as weighted AoI in this paper) is often used as the metric for quantifying the timeliness of communication. To minimize the weighted AoI, the source transmissions need to be scheduled appropriately. 

\emph{Maximum age first (MAF)} policy \cite{maf1,maf2,maf3}, where the source with highest instantaneous age is scheduled, and \emph{maximum-age-difference drop (MAD)} policy \cite{mad}, where the source which would result in the maximum drop in age is scheduled, have been extensively studied in the literature for this multi-source setting. Max-weight, Whittle-index policies \cite{age_aware1,age_aware2,age_aware3}, along with MAF and MAD are a few examples of age-aware scheduling policies which require the transmitter to continuously track the age of the sources and therefore can introduce a significant communication overhead, especially in channels susceptible to packet drops (or errors). Moreover, in open-loop communication systems where the feedback on packet drops is absent (i.e., only the channel service time is known), these age aware schemes are not feasible. Therefore, age-agnostic cyclic scheduling has recently become a viable solution to mitigate this communication overhead\cite{cyclic1,cyclic2,eywa,gau23}. 

\begin{figure}[t]
    \centering
    \includegraphics[scale=0.8]{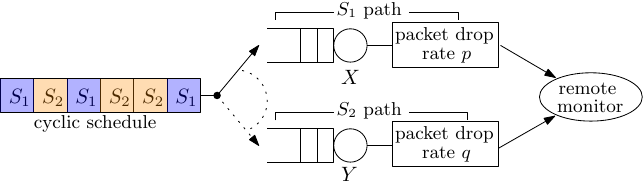}
    \caption{System model.}
    \label{fig:sys}
    \vspace*{-0.2cm}
\end{figure}

A framework named \emph{Eywa} was introduced in \cite{eywa} where the goal is to construct \emph{almost uniform cyclic schedulers (AUS)}, which is a special class of cyclic schedulers designed to distribute the scheduling instances of a given source as uniformly as possible within the cycle. \emph{Eywa} works in the discrete time setting and assumes that all the sources have the same deterministic service times with heterogeneous packet errors. However, in a more practical setting, the service times of different sources may be different, in which case reference \cite{gau23} obtains the optimal cyclic scheduler that minimizes the weighted AoI for two heterogeneous sources in the absence of packet errors. Reference \cite{gau23} shows that in the absence of packet errors, the optimal cyclic schedulers are of the form $(1,\Theta)$ which represents a cyclic schedule where one scheduling instance of one of the sources is followed by $\Theta$ scheduling instances of the other source. 

In this paper, we extend the work done in \cite{gau23} to the case of packet errors, and find the best cyclic scheduler which is resilient even in the presence of packet errors in the channel. We note that there are some similarities but significant key differences between the optimal cyclic scheduler constructed in this work and the one constructed in the absence of packet errors \cite{gau23}. In the presence of packet errors, we show that a near-optimal cyclic scheduler is a mixture of $(1,\Theta)$ and $(1,\Theta+1)$ cyclic schedulers. We emphasize the fact that even though we start off with cyclic schedulers, due to packet drops, the actual schedule of successful transmissions will no longer behave according to the constructed schedule. Hence, AoI computation is a difficult task even for two sources. 

To summarize our contributions:
\begin{itemize}
    \item We provide a Markov chain based formulation to compute the weighted AoI for two heterogeneous sources following a cyclic schedule in the presence of packet drops, which could also be extended to any number of sources.
    \item We provide an algorithm to produce cyclic schedulers which we prove to be near-optimal, i.e., given any $\epsilon>0$, we can find a cyclic schedule whose weighted AoI is within $\epsilon$ of the actual optimum.
\end{itemize}

\section{System Model}
Consider a communication system (shown in Fig.~\ref{fig:sys}) consisting of two heterogeneous sources which provide status updates to a remote monitor through a shared random delay (or random service time) channel with packet drops where the two sources are scheduled according to a cyclic pattern. Once the transmitter has finished transmitting the current sample, it will immediately sample from the next source in the schedule and start transmitting this new sample. The information on whether the transmitted packet is dropped at the channel or not, is not available to the transmitter as in \cite{eywa}.

Let the two sources be denoted as $S_1$ and $S_2$ with the channel service times of the two sources given by the random variables $X$ and $Y$  with means $s_1$, $s_2$ and variances $v_1$, $v_2$, respectively. Let the packet drop probability of the two sources be $p$ and $q$. Let us denote by $u$, the cycle length and by $u_1$, the number of instances of $S_1$ within the cycle. Then, $u_2=u-u_1$ is the number of scheduling instances for $S_2$. Let $\bm{r}=\{r_1,r_2,\dots,r_{u_1}\}$ represent the placement vector of the schedule with respect to $S_1$, where $r_i$ is the number of $S_2$ scheduling instances between the $i$th and the $(i+1)$th scheduling instances of $S_1$. Then, $(u,u_1,\bm{r})$ represents any cyclic schedule of the two sources uniquely. For example, the schedule $\{S_1,S_2,S_1,S_2,S_2\}$ will be represented by the tuple $(5,2,\{1,2\})$. Any feasible cycle should allocate at least one scheduling instance for each source. Therefore, for a feasible schedule, $u>u_1\geq1$.

The first step towards realizing our goal is to analytically obtain the mean AoI (or average AoI or AoI in short) of each of the two sources of the cyclic schedule by taking into account the packet drop probabilities.

\section{AoI Analysis}\label{sec:aoi_analysis}
\begin{figure}
    \centering
    \includegraphics[width=\columnwidth]{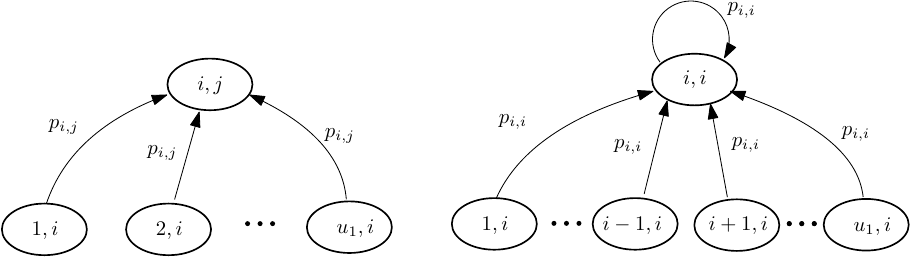}
    \caption{Partial state transition diagram for $u_1>2$.}
    \label{fig:u-MC}
    \vspace*{-0.4cm}
\end{figure}

To derive the AoI expression, we first consider $S_1$ and use a Markov chain formulation to characterize its AoI process. In one scheduling cycle, there are $u_1$ scheduling instances of $S_1$. Therefore, based on where two consecutive AoI drops occur relative to the cycle, we can define $u_1^2$ states for the Markov chain. Let $(i,j)$ for $i,j \in \{1,2, \ldots, u_1\}$ denote an AoI cycle starting from the $i$th scheduling instance of $S_1$ and ending in the $j$th scheduling instance of $S_1$. Note $(i,j)$ represents the $i$th and the $j$th scheduling instances of $S_1$ relative to one cycle of the schedule. Each state transition occurs after successfully (without packet drop) transmitting a  sample from $S_1$. Fig.~\ref{fig:u-MC} shows a partial state transition diagram of the Markov chain.

As shown in Fig.~\ref{fig:u-MC}, state $(i,j)$ is only directly accessible by states of the form $(k,i)$ and they all share the same transition probability $p_{i,j}$. Only states of the form $(i,i)$ have self transitions. The transition probabilities are given as,
\begin{align}
    p_{i,j} &= \frac{(1-p)p^{j-i-1}}{1-p^{u_1}}, & \text{if } j>i, \\
    p_{i,j} &= \frac{(1-p)p^{j+u_1-i-1}}{1-p^{u_1}}, & \text{if }  i\geq j.
\end{align}
Since these are finite-state Markov chains with no absorbing states for $p>0$, they are positive recurrent and irreducible. Since they contain self-loops, they are aperiodic and hence are ergodic. Let $\pi=\{\pi_{(i,j)}\}$  for $i,j \in \{1,2, \dots ,u_1\}$ denote the stationary distribution and $P_{(k,l),(i,j)}$ denote the transition probability from state $(k,l)$ to state $(i,j)$ of the Markov chain. Then, the stationary distribution takes the following form,
\begin{align}
    \pi_{i,j} = \frac{p_{i,j}}{u_1}.
\end{align}

To compute the average AoI, we partition the AoI (or age) graph based on the states of the Markov chain. Let $A_{i,j}$ denote the area under the age curve  when the two consecutive age drops are placed at the $i$th and $j$th scheduling instances of the cycle. Let $T_{i,j}$ denote the  time spent for the aforementioned age drops. Since the Markov chain is ergodic, the average AoI of $S_1$ denoted by $\e[\Delta_t^1]$, is given by,
\begin{align}
    \mathbb{E}[\Delta_t^1] = \frac{\sum_{i=1}^{u_1}\sum_{j=1}^{u_1}\pi_{i,j}\e[A_{i,j}]}{\sum_{i=1}^{u_1}\sum_{j=1}^{u_1}\pi_{i,j}\e[T_{i,j}]}. \label{eqn:aoi_stat}
\end{align}

\begin{figure}
    \centering
    \includegraphics[width=0.95\columnwidth]{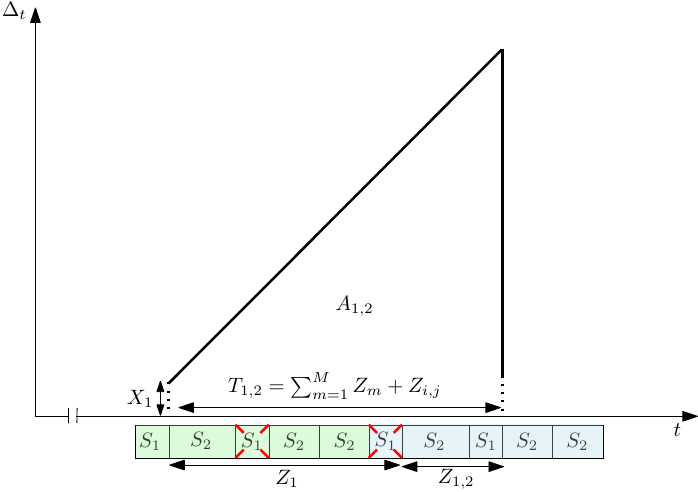}
    \caption{AoI graph of the schedule $\{S_1,S_2,S_1,S_2,S_2\}$ showing a realization of the area $A_{1,2}$ when $M = 1$.}
    \label{fig:aoi}
    \vspace*{-0.4cm}
\end{figure}

Therefore, to compute the average AoI, we need to find $\e[A_{i,j}]$ and $\e[T_{i,j}]$. $A_{i,j}$ consists of the segments of the age curve with consecutive age drops happening at the $i$th and $j$th scheduling instances of $S_1$. Suppose an AoI drop happens at $i$th scheduling instance, then the next AoI drop may happen at the $j$th scheduling instance after going through multiple rounds of the entire cycle. This is illustrated in Fig.~\ref{fig:aoi}, where failed $S_1$ transmissions are crossed out in red. Let $M$ denote the number of rounds of the entire cycle that have elapsed before the next successful transmission of a $S_1$ sample occurring at the $j$th scheduling instance. For example, if the next successful transmission occurs in the second round, then $M=1$. Then, $M=\hat{M}-1$, where $\hat{M} \sim \text{Geom}(p^{u_1})$. Let $Z_{i,j}$ denote the time duration elapsed starting from the beginning of the $(M+1)$th round to the AoI drop occurring at the $j$th scheduling instance of $S_1$. Let $Z_m$ denote the time duration of the $m$th round. Then, $\e[A_{i,j}]$ can be written as,
\begin{align}
    \!\mathbb{E}[A_{i,j}] &= \frac{\e\left[\left(2X+\sum\limits_{m=1}^MZ_m+Z_{i,j}\right)\left(\sum\limits_{m=1}^MZ_m+Z_{i,j}\right)\right]}{2} \nonumber\\
    &=\frac{\tilde{m}\hat{s}^2+\overline{m}(2s_1\hat{s}+\hat{v})+\tilde{z}_{i,j}+2\overline{z}_{i,j}(\overline{m}\hat{s}+s_1)}{2},\label{eqn:ai}
\end{align}
where $\hat{s}=u_1s_1+u_2s_2$, $\hat{v} = u_1v_1+u_2v_2$,  $\e[M] = \overline{m}$, $\e[M^2] = \tilde{m}$, $\e[Z_{i,j}] = \overline{z}_{i,j}$ and $\e[Z_{i,j}^2] = \tilde{z}_{i,j}$. Similarly, $\e[T_{i,j}]$ can be written as,
\begin{align}
    \e[T_{i,j}] =\e\left[\sum_{m=1}^MZ_m+Z_{i,j}\right] = \overline{m}\hat{s}+\overline{z}_{i,j} \label{eqn:T}
\end{align}
Then, by substituting (\ref{eqn:ai}) and (\ref{eqn:T}) into (\ref{eqn:aoi_stat}), we find the average AoI of $S_1$ as follows,
\begin{align}
    \e[\Delta_t^1]= &\frac{\tilde{m}\hat{s}^2+\overline{m}(2s_1\hat{s}+\hat{v})+\sum_{i,j}\pi_{i,j}\tilde{z}_{i,j}}{2(\overline{m}\hat{s}+\sum_{i,j}\pi_{i,j}\overline{z}_{i,j})}\nonumber\\
    &+\frac{(\overline{m}\hat{s}+s_1)\sum_{i,j}\pi_{i,j}\overline{z}_{i,j}}{\overline{m}\hat{s}+\sum_{i,j}\pi_{i,j}\overline{z}_{i,j}} \label{eqn:u1_aoi}
\end{align}

Let $a=\frac{u_2}{u_1}$, $s=as_2+s_1$ and $v=av_2+v_1$. Then, based on the symmetric nature of the stationary distribution, we can further simplify the above expression using the following,
\begin{align}
     \sum_{i,j}\pi_{i,j}\overline{z}_{i,j} &= s\frac{(1-p)}{(1-p^{u_1})}\sum_{i=1}^{u_1}ip^{i-1},\label{eqn:pi_z}\\
    \sum_{i,j}\pi_{i,j}\tilde{z}_{i,j} &=\frac{(1-p)}{(1-p^{u_1})}\sum_{i=1}^{u_1}\left[iv+i^2s^2\right]p^{i-1}\nonumber\\
    &\quad+s_2^2\frac{(1-p)}{u_1(1-p^{u_1})}\sum_{i=1}^{u_1}(\tilde{r}(i)-u_1a^2i^2)p^{i-1}.\label{eqn:pi_zz}
\end{align}
In  (\ref{eqn:pi_zz}), $\Tilde{r}(i)$ is defined as follows,
\begin{align}
    \Tilde{r}(i) = \sum_{j=1}^{u_1}\left(\sum_{k=j}^{j+i-1}\gamma_k\right)^2
\end{align}
where $\bm{\gamma}= \{r_1,r_2, \dots,r_{u_1},r_1,r_2, \dots,r_{u_1}\}$.
The $\tilde{r}(i)$ term is simply the sum of squared sum of $i$ consecutive elements in the placement vector. For example, $\tilde{r}(1)$, $\tilde{r}(2)$ and $\tilde{r}(3)$  are,
\begin{align}
    \tilde{r}(1) &= r_1^2 + r_2^2 + \dots +r_{u_1}^2\\
    \tilde{r}(2) &= (r_1\!+\! r_2)^2 + (r_2\!+\! r_3)^2 +\dots+(r_{u_1}\!+\!r_1)^2\\
    \tilde{r}(3) &= (r_1\!+\!r_2\!+\!r_3)^2 + (r_2\!+\!r_3\!+\!r_4)^2\dots + (r_{u_1}\!+\!r_1\!+\!r_2)^2
\end{align}
Substituting (\ref{eqn:pi_z}) and (\ref{eqn:pi_zz}) into (\ref{eqn:u1_aoi}) yields average AoI of $S_1$ as,
\begin{align}
    \e[\Delta_t^1] &= \frac{(1+p)}{2(1-p)}s + \frac{v}{2s}+s_1 \nonumber\\
    &\quad+s_2^2\frac{(1-p)^2}{2su_1(1-p^{u_1})}\sum_{i=1}^{u_1}(\tilde{r}(i)-u_1a^2i^2)p^{i-1}\label{eqn:aoi_1}
\end{align}
Similarly, the average AoI of $S_2$ denoted by $\e[\Delta_t^2]$, can be found based on the placement vector of $S_2$ (by reversing the roles of $S_1$ and $S_2$).

\begin{remark}
    When $p=0$, (\ref{eqn:aoi_1}) reduces to the AoI obtained in \cite{gau23} which is a special case of the model studied here. 
\end{remark}
\begin{remark}
    By appropriately modifying $Z_m$ and $Z_{i,j}$ in  (\ref{eqn:ai}), we can find the average AoI of a cyclic scheduler with arbitrary number of sources, i.e., $N>2$.
\end{remark}

\section{Minimizing the Weighted AoI }
Let $w_1$ and $w_2$ be the normalized weights associated with $S_1$ and $S_2$, respectively. Our goal is to find the best possible cyclic scheduler that minimizes the weighted AoI $\e[\Delta_t^w]$ of the sources given below,
\begin{align}
    \e[\Delta_t^w]=w_1\e[\Delta_t^1]+w_2\e[\Delta_t^2].
\end{align}
We start by analyzing the optimization of $\e[\Delta_t^1]$ alone, and later show that independently minimizing $\e[\Delta_t^1]$ and $\e[\Delta_t^2]$ jointly minimizes both, and hence minimizes $\e[\Delta_t^w]$. 

To proceed with the analysis, for fixed $u_1$ and $u_2$, we need to find the optimal placement vector $\bm{r}$ that minimizes (\ref{eqn:aoi_1}). By relaxing the integer constraint on the placement vector, it follows that the minimum is achieved when all $r_i$ are equal, i.e., $r_i=a$. Subsequently applying the integer constraints yields that $r_i$ is either $\ceil{a}$ or $\lfloor a\rfloor$. The structure of the optimal placement vector is then,
\begin{align}
    r_i = 
    \begin{cases}
        \lfloor a\rfloor, &  \# u_1(\ceil{a}-a),\\
        \ceil{a},  &  \# u_1(1+a-\ceil{a}).
    \end{cases}\label{eqn:r_i}
\end{align}
where $\#$ denotes the number of elements of each term. 

Now that we know the structure of the optimal placement vector, next we need to find the optimal arrangement of $\ceil{a}$ and $\lfloor a\rfloor$ terms within the placement vector. This is one of the key differences from the work in \cite{gau23} where the ordering of the placement vector is inconsequential. Note that to minimize $\tilde{r}(i)$ we need to spread the elements such that they are as uniform as possible within every window of consecutive $i$ terms. This is achieved by hierarchically spreading different sub-blocks of the placement vector as given by Algorithm~\ref{alg:place}.

\begin{algorithm}[tbh]
\caption{Uniform arrangement of placement vectors}\label{alg:place}
\begin{algorithmic}
\Require $u_1$,$u_2$
\State $a=\frac{u_2}{u_1}$, $b_1=\lfloor a \rfloor$, $b_2=\ceil{a}$
\State $c_1=u_1(\ceil{a}-a)$, $c_2=  u_1(1+a-\ceil{a})$
\While{$\min(c_1,c_2)>1$}
    \If {$c_1>c_2$}
        \State $c_1 \gets c_2$, $c_2 \gets c_1$
        \State $b_1 \gets b_2$, $b_2 \gets b_1$
    \EndIf
    \State $c = \frac{c_2}{c_1}$
    \State $b_1 \gets \{b_1,b_2 \times \lfloor c \rfloor\}$ , $b_2 \gets \{b_1,b_2 \times \ceil{c}\}$
    \State $c_1 \gets c_1(\ceil{c}-c)$, $c_2 \gets c_1(1+c-\ceil{c})$
\EndWhile
\end{algorithmic}
\end{algorithm}
In Algorithm~\ref{alg:place}, $b_2 \gets \{b_1,b_2 \times \ceil{c}\}$ implies that the new $b_1$ is an array consisting of one instance of $b_1$ and $\ceil{c}$ instances of $b_2$. As an application of Algorithm~\ref{alg:place}, let us consider a scenario with $u_2 = 41$ and $u_1=11$, hence $u=52$. One possible placement vector for this case is $\bm{r}=\{3,3,3,4,4,4,4,4,4,4,4\}$ which is a vector of $3$ threes and $8$ fours. Now, we need to uniformly arrange this placement vector. For this purpose, we try to uniformly distribute the $8$ fours among the $3$ threes. This will give us two blocks of $\{3,4,4,4\}$ and one block of $\{3,4,4\}$. Next, we need to spread these sub-blocks as uniformly as possible. Since the minimum of the number of instances of these two blocks is one, the algorithm stops at this stage and returns $r=\{3,4,4,4,3,4,4,4,3,4,4\}$ as the optimal placement vector. The optimal placement vector generated by Algorithm~\ref{alg:place} is always a mixture of placement vectors of the form $r=\{\Theta\}$ and $r=\{\Theta+1\}$, where $\Theta = 3$ in this example.

Note that the above treatment tries to minimize the average AoI of $S_1$ by scheduling it as uniformly as possible for a fixed $u_1$ and $u_2$. When we uniformly distribute $S_1$ transmissions, we also notice that at the same time it allows us to uniformly distribute the $S_2$ as well. Hence, the above structure of the placement vector jointly minimizes the average AoI of both $S_1$ and $S_2$. This is one of the similarities with the work in \cite{gau23}. Therefore, what remains is to find the best possible $u_1$ and $u_2$. This can be reduced to finding the best possible rational number for $a$ that minimizes the weighted AoI. 

To find the optimal $a$, we first find the average AoI of the round robin (RR) policy, denoted by $\e[\Delta_{RR}]$, which can be obtained by setting  $u_1 = u_2 = a = 1$ in (\ref{eqn:aoi_1}).
Since the first term in $\e[\Delta_t^1]$ is linear in $a$, we only have to consider values of $a$ for which the term  $w_1\e[\Delta_t^1]$ is less than the weighted AoI of the RR policy. Let the $a_{max}$ be the smallest $a$ such that $w_1\e[\Delta_t^1]> \e[\Delta_{RR}]$. Similarly, we can find a $a_{min}$ value by considering  $w_2\e[\Delta_t^2]$. Hence, the search space of $a$ is bounded by $a_{min}$ and $a_{max}$. 

From this point onwards, we represent a cyclic schedule only using the tuple $(u_1,u_2)$ where the optimal placement vector is found using Algorithm~\ref{alg:place}. For a fixed $a$, consider the patterns  $(u_1,u_2)$ and $(ku_1,ku_2)$  where $k\in \mathbb{N}$. If $\bm{r}$ is the optimal placement vector for $(u_1,u_2)$, by simply repeating $\bm{r}$, $k$ times, we can obtain the optimal placement vector for the pattern $(ku_1,ku_2)$ which yields the same average AoI. Therefore, in the bounded region for $a$, we only need to compare rationals in their simplest form. Algorithm~\ref{alg:optim} can be used to find a near-optimal $a$ and $\bm{r}$ which minimize the weighted AoI. In Algorithm~\ref{alg:optim}, $\e[\Delta_i^r]$ represents the AoI of the $i$th source with respect to the placement vector of $S_1$, $\mathbf{ALG1}(\tilde{u}_1,\tilde{u}_2)$ is the output of Algorithm~\ref{alg:place} for the selected $\tilde{u}_1$, $\tilde{u}_2$ and $\mathbf{Coprime}(u_1,u_2)$ returns the co-primes of $u_1$ and $u_2$. Even though the average AoI depends on both $a$ and $u_1$, we show in the next section that fixing $u_1$ to a large value and finding $a$ using Algorithm~\ref{alg:optim} is sufficient to be as close as desired to the optimal schedule.

\begin{algorithm}[t]
    \caption{Algorithm to find the near-optimal cyclic pattern}\label{alg:optim}
    \begin{algorithmic}
        \Require $\alpha$ sufficiently large integer.
        \State $u_1= \alpha$, $u_2=\alpha$, $AoI_{min}=\e[\Delta_{RR}]$, $r^*=\{1\}$
        \State $a_{max} = \inf\{a:w_1\e[\Delta_1]>\e[\Delta_{RR}]\}$
        \While{$\frac{u_2}{u_1}<a_{max}$}
            \State $(\tilde{u}_1,\tilde{u}_2) = \mathbf{Coprime}(u_1,u_2)$ 
            \State $r= \mathbf{ALG1}(\tilde{u}_1,\tilde{u}_2)$ 
            \State $AoI = w_1\e[\Delta_1^r]+w_2\e[\Delta_2^r]$
            \If {$AoI_{min}> AoI$}
                \State $r^*=r$ , $AoI_{min}=AoI$
            \EndIf
            \State $u_2=u_2+1$
        \EndWhile
        \State $u_1= \alpha$, $u_2=\alpha$
        \State $a_{min} = \sup\{a:w_2\e[\Delta_2]>\e[\Delta_{RR}]\}$
        \While{$\frac{u_2}{u_1}>a_{min}$}
            \State $(\tilde{u}_1,\tilde{u}_2) = \mathbf{Coprime}(u_1,u_2)$  
            \State $r= \mathbf{ALG1}(\tilde{u}_1,\tilde{u}_2)$ 
            \State $AoI = w_1\e[\Delta_1^r]+w_2\e[\Delta_2^r]$
            \If {$AoI_{min}> AoI$}
                \State $r^*=r$, $AoI_{min}=AoI$
            \EndIf
            \State $u_1=u_1+1$
        \EndWhile
    \State{\textbf{Output:}} $r^*$
    \end{algorithmic}
\end{algorithm}

\section{Optimality of the Proposed Algorithm}
In this section, we will prove that for all $\epsilon>0$, the minimum weighted AoI obtained by the solution of Algorithm~\ref{alg:optim} can be made to be within $\epsilon$ of the actual optimum by choosing $\alpha$ in Algorithm~\ref{alg:optim} sufficiently large. Without loss of generality, assume optimal $a^*>1$. Let us analyze the age expression of $S_1$. Define the functions $f(a)$, $g(a)$, and $h(a)$ as follows,
\begin{align}
    f(a)& = \frac{(1+p)}{2(1-p)}s + \frac{v}{2s}+s_1,\\
    g(a) &= a^2s_2^2\frac{(1-p)^2}{2s(1-p^{u_1})}\sum_{i=1}^{u_1}i^2p^{i-1},\\
    h(a) &= s_2^2\frac{(1-p)^2}{2su_1(1-p^{u_1})}\sum_{i=1}^{u_1}\tilde{r}(i)p^{i-1}.
\end{align}
Then, $\e[\Delta_t^1] = f(a)-g(a)+h(a)$. In Algorithm~\ref{alg:optim}, we fix $u_1$ and increase $u_2$ since $a^*>1$. Since $f(a)$ is a continuous function of $a$, if $a$ is sufficiently close to $a^*$, $f(a) \approx f(a^*)$. For sufficiently large $u_1$, $\sum_{i=1}^{u_1}i^2p^{i-1}/(1-p^{u_1})$ is approximately a constant. Therefore, when $a$ is sufficiently close to $a^*$, $g(a)\approx g(a^*)$. Consider an $\alpha_0 \in \mathbb{N}$ such that $a_0=\frac{\alpha_0}{\alpha}<a^*<\frac{\alpha_0+1}{\alpha}=a_1$. If $\alpha$ is sufficiently large, for all $a$ such that $a_0<a<a_1$, $\ceil{a}=\ceil{a^*}$ and $\lfloor a \rfloor = \lfloor a^* \rfloor$. Thus, the placement vector for any $a \in [a_0,a_1]$ consists of either $\ceil{a^*}$ or $\lfloor a^* \rfloor$. 

Now, consider $a = (a_0+a_1)/2 = (2\alpha_0+1)/2\alpha$. The average AoI when $(u_1,u_2) = (\alpha,\alpha_0)$ is equal to the average AoI when $(u_1,u_2)= (2\alpha,2\alpha_0)$. The placement vector of $(2\alpha,2\alpha_0+1)$ would only differ in one term with the placement vector of $(2\alpha_0,2\alpha)$, where it is increased by 1.  Therefore, the difference in $\tilde{r}(i)$ of the two patterns will be bounded by $2(\ceil{a^*}+1)i^2$. Therefore, the difference in the average AoI is bounded by $2C(\ceil{a^*}+1)\sum_{i=1}^{u_1}i^2p^{i-1}/u_1$, where $C$ is a constant. Since $\alpha$ is large, we have made the assumption that $(1-p^{u_1}) \approx 1$. Therefore, by approaching $a^*$ in a bisection search, we can show that the AoI difference when $a=a^*$ and when $a$ equals one of the end points $a_0$ or $a_1$, will be bounded by a constant times $1/\alpha$. A similar argument holds for $\e[\Delta_t^2]$. Therefore, by selecting a large enough $\alpha$, we can make our solution as close as desired to the optimal solution. 

\begin{figure}[t]
    \centering
    \begin{subfigure}[b]{0.49\columnwidth}
         \centering
         \includegraphics[scale=0.3]{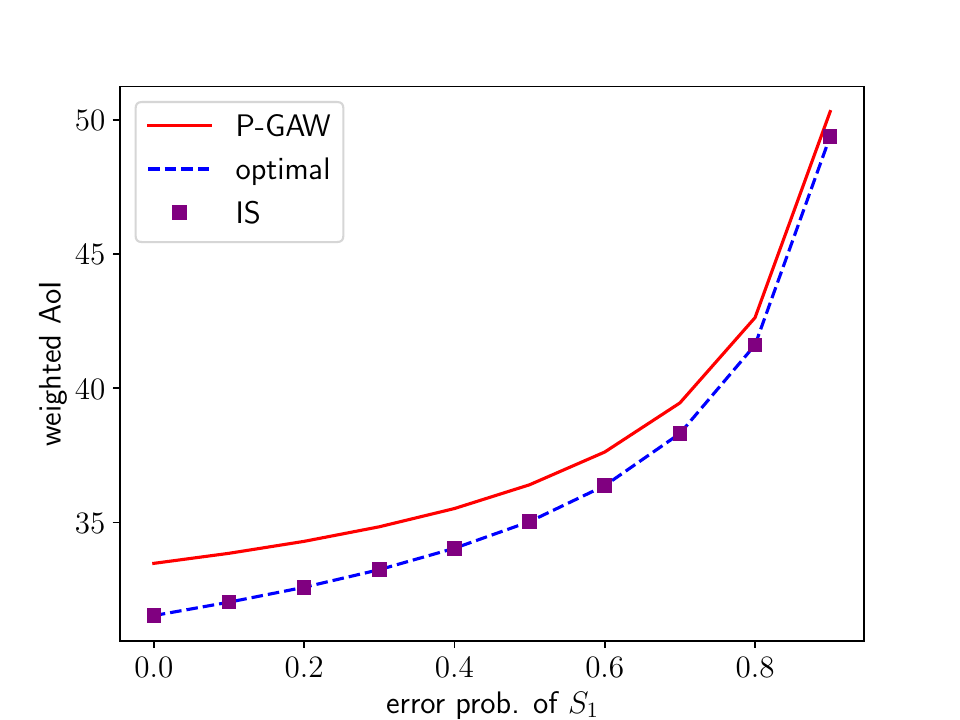}
         \caption{$s_1=2$}
     \end{subfigure}
    \begin{subfigure}[b]{0.49\columnwidth}
         \centering
         \includegraphics[scale=0.3]{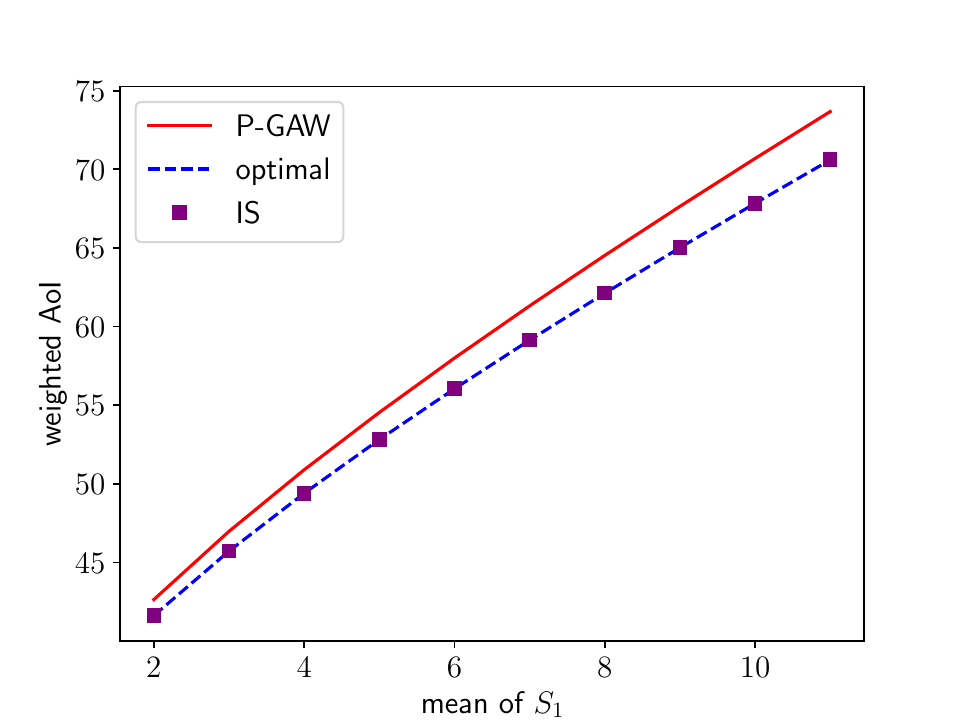}
         \caption{$p=0.8$}
     \end{subfigure}
    \caption{Variation of weighted AoI with packet drop probability and mean of $S_1$ for exponential service times ($s_2=3$, $w_1=0.2$, $w_2= 0.8$, $q=0.9$).}
    \label{fig:exp_var}
\end{figure}

\section{Numerical Results}
We compare the performance of our scheduler marked \emph{optimal} with the \emph{probabilistic generate-at-will} (P-GAW) \cite{gau23} and the \emph{Eywa}  \cite{eywa}. In P-GAW, the $i$th source is scheduled for transmission with probability $p_i$, with $p_1+p_2=1$ and the optimal $p_1$ values is found through a 1-dim exhaustive search.
 
\begin{figure}[t]
    \centering
    \includegraphics[scale=0.4]{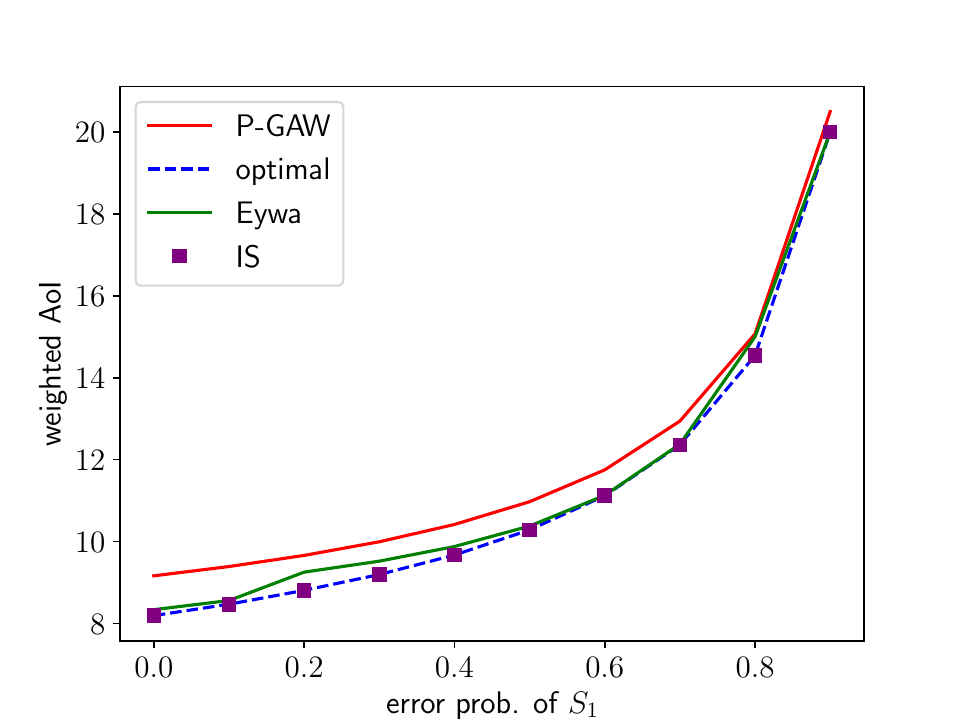}
    \caption{Variation of weighted AoI with packet drop probability of $S_1$ for deterministic service times ($s_1= s_2 =1$, $w_1=0.5$, $w_2= 0.5$, $q=0.9$).} 
    \label{fig:pgaw-eywa}
     \vspace*{-0.2cm}
\end{figure}

\begin{figure}[t]
    \centering
    \includegraphics[scale=0.4]{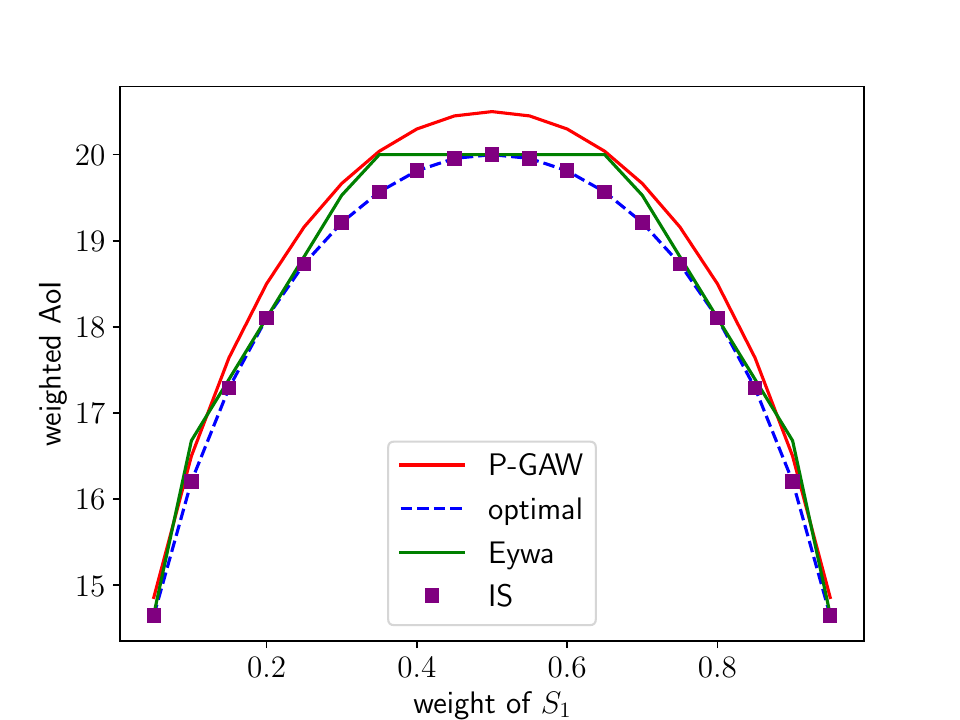}
    \caption{Variation of weighted AoI with weight of $S_1$ for deterministic service times ($s_1= s_2 =1$, $p=0.9$, $q=0.9$).} 
    \label{fig:eywa_vs_w}
    \vspace*{-0.2cm}
\end{figure}

We first compare the performance of our scheduler against the P-GAW scheduler for exponential channel service times. For this, we fix the parameters of $S_2$ and vary either the packet drop probability of $S_1$ or the mean of $S_1$. As seen in Fig.~\ref{fig:exp_var},  our scheduler achieves significantly lower average AoI than P-GAW. This is to be expected since the average AoI of a P-GAW model is simply an expectation of all deterministic cyclic schedules and ours is the closest to the best cyclic schedule. For identical deterministic channel service times (i.e., $s_1=s_2$ and $v_1=v_2=0$), we compare the performance of our scheduler against both P-GAW and \emph{Eywa}. We fix the parameters of $S_2$ and in one experiment we vary the packet drop probability of $S_1$, and in the other, we vary the weight given to the two sources. As shown in figures~\ref{fig:pgaw-eywa} and \ref{fig:eywa_vs_w}, our scheduler outperforms both \emph{Eywa} and P-GAW.

In each of the experiments, we also adopt the insertion search (IS) algorithm introduced in \cite{gau23} as a heuristic-based method for the construction  of cyclic schedulers for more than two sources in the absence of packet errors. IS algorithm starts off from a RR pattern and constructs a cyclic schedule in an iterative fashion. In each iteration, the source that will result in the lowest average AoI is selected by considering all possible sources, and all possible locations in the current pattern where a new scheduling instance can be inserted. \cite{gau23} shows that IS is optimal for two sources in the absence of packet errors. In the presence of packet errors, our experimental results show that the IS algorithm performs very close to the optimum solution. However, IS has a higher computational complexity than our algorithm and is not proven to be optimal in the presence of packet errors.

\section{Conclusion}
In this work, we have presented a theoretical framework to compute the average per-source AoI for cyclic scheduling in the presence of packet errors for two heterogeneous sources, and we provided several algorithms which enable us to find a near-optimal cyclic scheduler to minimize the weighted AoI. We have shown through numerical results and simulations that our scheduler outperforms existing age-agnostic schedulers.

\bibliographystyle{unsrt}
\bibliography{refs}

\end{document}